\documentclass[aps,prl,preprint]{revtex4-1}

\usepackage{natbib}
\usepackage[T1]{fontenc}
\usepackage{rotating}
\usepackage{bm}        % \boldsymbol (bold math)
\usepackage{amssymb}   % \mathbb     (blackboard bold)
\usepackage{float}
\usepackage{diagbox}
\usepackage{braket}
\usepackage{color} 
\usepackage{colortbl}
\usepackage{amsmath}

\usepackage[ruled,vlined]{algorithm2e}
\usepackage{listings}
\usepackage{indentfirst}
\usepackage{blkarray}

\usepackage{dcolumn}% Align table columns on decimal point
\usepackage[left=30mm,top=30mm,right=30mm,bottom=30mm]{geometry}
\usepackage{etoolbox} %required for cover page
\usepackage{booktabs}

\usepackage[T1]{fontenc}
\usepackage[utf8]{inputenc}
\usepackage{bm}
\usepackage{subcaption}
\usepackage{amsmath}
\usepackage{amsfonts}
\usepackage{mathtools}
\usepackage{xcolor}
\usepackage{float}
\usepackage{hyperref}
\usepackage[capitalise]{cleveref}
\usepackage{enumitem,kantlipsum}
\usepackage{amssymb}
\usepackage{jabbrv}

\makeatletter

    \def\CT@@do@color{%
      \global\let\CT@do@color\relax
            \@tempdima\wd\z@
            \advance\@tempdima\@tempdimb
            \advance\@tempdima\@tempdimc
    \advance\@tempdimb\tabcolsep
    \advance\@tempdimc\tabcolsep
    \advance\@tempdima2\tabcolsep
            \kern-\@tempdimb
            \leaders\vrule
                    \hskip\@tempdima\@plus  1fill
            \kern-\@tempdimc
            \hskip-\wd\z@ \@plus -1fill }
    \makeatother

\def\k1{k_1}
\def\k2{k_2}
\def\q1{q_1}
\def\q2{q_2}

\def\({\left (}
\def\){\right )}
\def\[{\left [}
\def\]{\right ]}

\newcommand{\beq}{\begin{equation}}
\newcommand{\eeq}{\end{equation}}

\DeclareMathAlphabet\mathbfcal{OMS}{cmsy}{b}{n}

\begin{document}

\date{\today}
\flushbottom \draft

\title{Efficient interpolation of molecular properties across chemical compound space with low-dimensional descriptors}% Force line breaks with \\

\author{Yun-Wen Mao}
 %\altaffiliation{Department of Chemistry, University of British Columbia, Vancouver, B.C. V6T 1Z1, Canada \\
%Stewart Blusson Quantum Matter Institute, Vancouver, B.C. V6T 1Z4, Canada}%Lines break automatically or can be forced with \\
\author{Roman V. Krems}%
% \email{rkrems@chem.ubc.ca}
\affiliation{%
Department of Chemistry, University of British Columbia, Vancouver, B.C. V6T 1Z1, Canada \\
Stewart Blusson Quantum Matter Institute, Vancouver, B.C. V6T 1Z4, Canada
}%

\date{\today}% It is always \today, today,

\begin{abstract}
We demonstrate accurate data-starved models of molecular properties for interpolation in chemical compound spaces with low-dimensional descriptors.
 Our starting point is based on three-dimensional, universal, physical descriptors derived from the properties of the distributions of the eigenvalues of Coulomb matrices.  To account for the shape and composition of molecules, we combine these descriptors with six-dimensional features informed by the Gershgorin circle theorem.  We use the nine-dimensional descriptors thus obtained for Gaussian process regression based on kernels with variable functional form, leading to extremely efficient, low-dimensional interpolation models. The resulting models trained with 100 molecules are able to predict the product of entropy and temperature ($S \times T$) and zero point vibrational energy (ZPVE) with the absolute error under 1 kcal mol$^{-1}$ for $> 78$ \%  and under 1.3  kcal mol$^{-1}$ for $> 92$ \% of molecules in the test data. The test data comprises 20,000 molecules with complexity varying from three atoms to 29 atoms and the ranges of $S \times T$ and ZPVE covering  36  kcal mol$^{-1}$ and 161  kcal mol$^{-1}$, respectively.   
 We also illustrate that the descriptors based on the Gershgorin circle theorem yield more accurate models of molecular entropy than those based on graph neural networks that explicitly account for the atomic connectivity of molecules. 
  
\end{abstract}

%Use showkeys class option if keyword
\maketitle
\section{\label{sec:level1} Introduction}

The number of molecules with properties suitable for medicinal applications is estimated to be between $10^{23}$ and $10^{60}$ \cite{polishchuk2013estimation}. 
This large number underscores the importance of the development of efficient methods for predicting molecular properties for any given combination of chemically bonded atoms. 
As demonstrated by recent work \cite{Weinreich2021, hansen2015}, prediction of molecular properties can be achieved by interpolation in chemical compound space, defined as an abstract space with individual molecules as elements of the space, by machine learning (ML) models.   
A chemical space is often, although not always, restricted to a particular number and combination of atoms (for example, a space of molecules with chemical formula C$_7$H$_{10}$O$_2$). 
Interpolation of molecular properties across chemical space meets with four main challenges: (i) the vast number of molecules; (ii) the requirement to represent molecules by numerical descriptors that are universal, efficient and accurate; (iii) the sensitivity of molecular properties to small structural changes in atomic arrangements; (iv) lack of structure in chemical compound space. While interpolation of molecular properties in chemical spaces is often considered as a big data problem, requiring information from large sets of molecules, the focus of many recent studies has been on data-efficient models \cite{Browning2017, Gubaev2018, FCHL19, Bogojeski2020}. %data-efficient-models
The efficiency of ML predictions of molecular properties can be enhanced by building underlying symmetries into ML models \cite{Gastegger2018, Rostami2018, Qiao2020}, by improving the molecular descriptors \cite{musil2021physics, Gallegos2021, Willatt_2018, Langer2022, FCHL19, Faber2017, Faber2018} and by improving the ML models through kernel optimization for kernel models \cite{duvenaud2013structure, duvenaud2011additive, Deng2020} or optimization of neural network (NN) architecture and algorithm hyperparameters for NN-based models \cite{Gilmer2017, schnet_2018, Smith2019}.

Key to any ML predictions of molecular properties are molecular descriptors that serve as inputs into supervised learning models. The molecular descriptors must be numerical, universal (i.e. describe any molecule), unique, 
and embody all of the relevant features, such as the shape of molecules, the arrangement of chemical bonds, bond strengths and bond order, vibrational frequencies and conformational degrees of freedom. 
Many variations of molecular descriptors have recently been proposed, including Coulomb matrices \cite{PRL_CM_2011}, bag of bonds (BoB) features \cite{BoB_2015}, smooth overlap of atomic positions (SOAP) \cite{SOAP_2012}, Faber-Christensen-Huang-Lilienfeld (FCHL19) representation \cite{FCHL19}. 
Another possibility is to represent molecules by a Graph Neural Network (GNN) \cite{kipf2016semi,Wu2019} that explicitly accounts for the spatial arrangement and connectivity of atoms in a molecule. 
GNN can be combined with other features embodying molecular properties, yielding accurate molecular descriptors \cite{Zhu2022,Yang2019,9338432, Qiao2020}. 
However, for complex molecules, these descriptors are necessarily high-dimensional. While high-dimensional descriptors can be used as inputs into isotropic ML models (for example ML models with kernels that do not discriminate between different input dimensions) \cite{williams2006gaussian}, such models do not account for inherent differences in scales of different features. Willatt et. al. \cite{Willatt_2018} showed that reducing the dimensionality of descriptors may improve the accuracy of ML predictions. In addition, Musil et. al. \cite{musil2021physics} argued that descriptors reflecting some fundamental principles yield more robust, transferable and data-efficient ML models. 
In addition to efficiency, low-dimensional descriptors enable the possibility of Bayesian optimization of molecular properties in chemical compound space. 
The present work demonstrates that it is possible to design low-dimensional descriptors of large polyatomic molecules for extremely efficient interpolation of entropy and zero point vibrational energy in chemical compound space.

Different molecular properties exhibit vastly different variation across chemical compound spaces. Therefore, it can be argued that the most data-efficient predictions must be based on models and descriptors tailored for the specific molecular property. 
In the present work, we test this conjecture by first considering interpolation of molecular entropy across chemical compound spaces and developing low-dimensional descriptors tailored for the prediction of entropy. 
Our starting point is based on three-dimensional, universal, physical descriptors derived from the properties of the distributions of the eigenvalues of the Coulomb matrix. 
We show that the largest eigenvalue and the two leading moments of the eigenvalue distribution provide the physical descriptors for molecular entropy, as  elucidated using the Gershgorin circle theorem \cite{gershgorin1931uber, saad2011numerical_gershgorin}. 
We illustrate that the accuracy of these descriptors can be enhanced by combining them with either: (1) the leading principle components of the output of GNN tailored for the prediction of entropy; or (2) with 
descriptors based on the Gershgorin circle theorem. We show that the models of entropy using descriptors based on the Gershgorin circle theorem perform better than the models using descriptors based on GNN. 

 By using the resulting nine-dimensional descriptors in Gaussian process regression with kernels optimized for pattern recognition \cite{duvenaud2013structure, duvenaud2011additive}, we build ML models capable of predicting 
entropy of 133,000 organic molecules with up to 29 atoms (HCNOF) using as few as 100 molecules for training. The predictions have chemical accuracy (with predicted entropy $\times$ temperature under 1 kcal mol$^{-1}$), illustrating, to the best of our knowledge, the most data-efficient models of entropy to date. We then illustrate that these descriptors, even though tailored for models of entropy, can be used to interpolate zero point vibrational energy (ZPVE). The variation of ZPVE in the chemical space considered here is much greater than the scale of values covered by entropy. However, we show that the nine-dimensional descriptors can be used to produce extremely data-starved yet accurate models of ZPVE, provided the kernels of the underlying models are properly optimized.  We illustrated that models trained by only 100 molecules produce ZPVE with mean absolute error {\it and} absolute error for $> 92$ \% of predictions under 1.3 kcal mol$^{-1}$, in  a chemical compound space including molecules with up to 29 atoms and covering the ZPVE range of 161 kcal mol$^{-1}$

\section{Dataset summary}
In this work, we focus on molecules from the QM9 dataset \cite{qm9_dataset}, where a chemical subspace consists of 133,885 stable organic molecules made up of a combination of the following atoms: CHONF. Within the QM9 dataset, there are 526 subsets of constitutional isomers, where constitutional isomers are molecules with the same chemical formula but different atom connectivity. Each subset includes at least 2 molecules, while the largest constitutional isomer subset includes 6059 molecules.

We reserve 20 $\%$ of these molecules for the hold-out test set, while samples from the remaining distributions of molecules are used to train the ML models.  
Since molecules in this chemical subspace range from 3 to 29 atoms in size,  we bin all molecules according to their size before randomly selecting the distributions for the training and test sets. 
This ensures the ensuing ML models sample the entire chemical subspace, both at the training and test stage.  Furthermore,  20 $\%$ of the training dataset is used as a model validation set. 
The validation set is used for optimizing the hyperparameters of the ML models.

The prediction accuracy of the ML models is quantified, for both the test and validation sets, by the root-mean-square error (RMSE)
\begin{equation}\label{RMSE_equation}
{\rm RMSE} = \sqrt{\frac{\Sigma_{i=1}^{N}(y_{i}-\hat{y_{i}})^2}{N}}
\end{equation}
or mean absolute error (MAE)
\begin{equation}\label{RMSE_equation}
{\rm MAE} = \frac{1}{N}\sum_{i=1}^{N} |y_i - \hat{y_i}|
\end{equation}
 evaluated over the number of samples $N$ in the test or validation set, as appropriate. Here, 
 $y_{i}$ and $\hat{y_{i}}$ are the observed entropy and ML predicted entropy of the $i^{th}$ molecule in the test/validation set, comprising $N$ molecules. Lower RMSE and MAE values indicate better accuracy of ML predictions.

\section{Method}

The main goal of the present work is to develop low-dimensional, general descriptors of molecules that can be applied to both small and large molecules and used for efficient interpolation of molecular properties across chemical compound spaces with a great variety of molecules.  We build these descriptors incrementally, by starting with the properties of distributions describing the Coulomb matrices \cite{PRL_CM_2011}, considering the enhancement of the descriptors by Graph Neural Networks, and using physical insights offered by the Gershgorin circle theorem \cite{Bellman+2021, saad2011numerical_gershgorin}. Once identified, the best low-dimensional descriptors are used as features of inputs into Gaussian process models with kernels optimized for a specific molecular property. The efficiency of the final models presented here is thus determined by both the efficiency of the molecular descriptors and the alignment of model kernels with the prediction targets. 

\subsection{Descriptor I: Distributions of eigenvalues of Coulomb matrices}
Coulomb matrix is a molecular descriptor developed by Rupp et al. \cite{PRL_CM_2011}. Each molecule is represented by a $n_{\rm max}\times n_{\rm max}$ symmetric matrix, where $n_{\rm max}$ is the number of atoms in the largest molecule in the chemical subspace. 
Each matrix element $M_{ij}$ in matrix $M$ is defined as
\begin{equation}\label{coulomb_mat}
  M_{ij} =
    \begin{cases}
      0.5Z_{i}^{2.4} & \text{for $i=j$}\\
      \frac{Z_i Z_j}{R_{ij}} & \text{for $i\neq j$}
    \end{cases}      
\end{equation}
where $Z_i$ and $Z_j$ are the atomic numbers of atoms $i$ and $j$, respectively, and $R_{ij}$ is the separation between atoms $i$ and $j$. The diagonal elements of the Coulomb matrix represent the polynomial fit relating the atomic number to the total energies of the free atoms \cite{PRL_CM_2011}. The off-diagonal elements describe the electrostatic interaction between atoms $i$ and $j$. If the size of a molecule is smaller than $n_{\rm max}$,  the matrix elements with $i$ or $j > n_{\rm max}$ are set to zero. The Coulomb matrix is invariant to molecular translations and rotations, but not invariant to permutations of the atoms. This problem can be overcome by diagonalizing the Coulomb matrix \cite{PRL_CM_2011,schrier2020can}. 

In the present work, we begin by implementing the basic three-dimensional (3D) descriptors of molecules using the largest eigenvalue $\epsilon_{\rm max}$, the mean $\mu(\epsilon > 1)$ and the standard deviation $\sigma(\epsilon > 1)$ of the distribution of the eigenvalues  of the Coulomb matrix ($\epsilon$) with values greater than 1. These 3D descriptors are used as inputs into the most basic ML models of entropy in this work. To enhance the expressiveness of the molecular descriptors and the accuracy of the ensuing models, we supplement these 3D vectors with features based on either the outputs of Graph Neural Networks, optimized to predict entropy as described in Descriptor II, or features based on the Gershgorin circle theorem as described in  Descriptor III.

\subsection{Descriptor II: Graph Convolution Networks} \label{gnn-section}

\begin{figure} [H]
\centering
\includegraphics[width=1\textwidth]{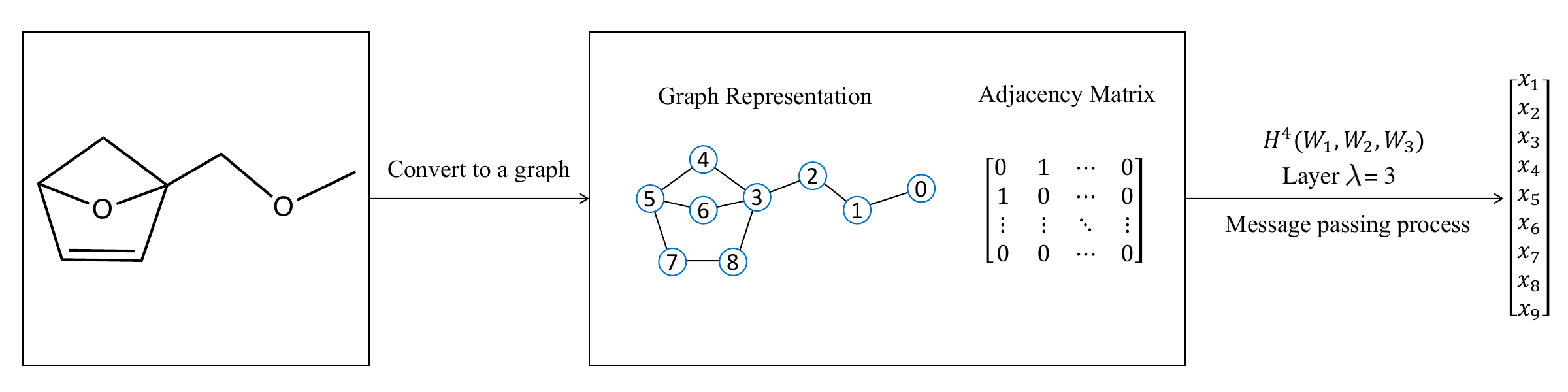}
\caption{\label{fig:Fig1} Schematic diagram illustrating the use of GCN to create molecular descriptor ($H^4_{\rm{C_2H_4O}}$) with C$_2$H$_4$O as an example. The GCN used in this work, as demonstrated in this figure, has four layers ($\lambda +1 = 4$). The resulting molecular descriptor is nine dimensional, because the largest molecule in QM9 dataset considered here has 9 heavy atoms.
}
\end{figure}

\begin{figure} [H]
\centering
\includegraphics[width=1\textwidth]{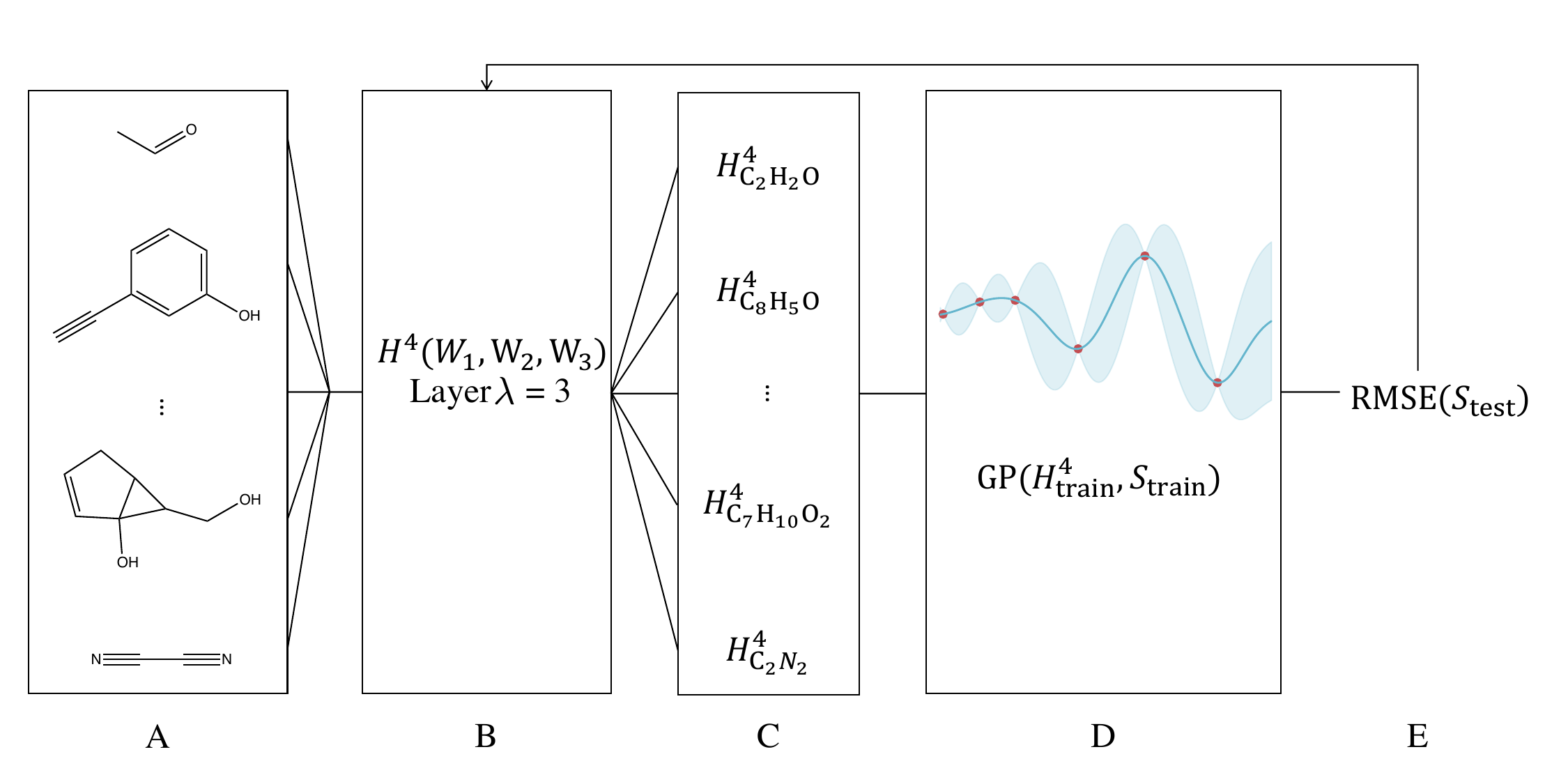}
\caption{\label{fig:Fig2} Schematic diagram of the algorithm used in the present work for training GCN by optimizing weight matrices ($W_1$, $W_2$, $W_3$). Each molecule in A is processed as described in Fig. \ref{fig:Fig1} to yield a unique GCN vector ($H^4_{\rm{molecule}}$), used as the molecular descriptor. A subset of molecules in A, yielding a set of vectors depicted in C, is used to train
a Gaussian process  (GP) model  of entropy. The RMSE of the GP predicted values is used in step E as the loss function to optimize the weight matrices ($W_1$, $W_2$, $W_3$) of GCN by gradient descent. 
}
\end{figure}

To enhance the accuracy of the resulting ML models as well as to explore the role of explicit graph descriptors for interpolation of molecular properties, we supplement the 3D descriptors derived from the Coulomb matrix by  descriptors based on the outputs of graph convolution networks (GCN). 
A molecule can be mapped onto a graph $G=(\nu$, $e$$)$, where the atoms and bonds are treated as nodes ($\nu$) and edges ($e$), respectively. Each graph is represented by an adjacency matrix $ A$ with matrix elements defined as
\begin{equation}\label{adj_mat_def}
  A_{ij} =
    \begin{cases}
      1 & \text{if $e_{ij} \in \varepsilon$}\\
      0 & \text{otherwise,}
    \end{cases}      
\end{equation}
where $e_{ij}$ is the edge between the $i^{th}$ and $j^{th}$ node. With the adjacency matrix thus defined, the $(\lambda+1)^{\rm th}$ layer of a multi-layer GCN is implemented using
\begin{equation}\label{gcn_l+1}
    H^{\lambda+1} = \phi (\hat{D}^{-\frac{1}{2}} \hat{A} \hat{D}^{-\frac{1}{2}} H^\lambda W^\lambda),
\end{equation}
where $\hat{D}$ is the diagonal matrix with elements
\begin{equation}\label{degree_mat}
    \hat{D}_{ii}=\sum_{j-1}^{m_{\rm max}}\hat{A}_{ij},
\end{equation}
$\hat{A} = A + I_{m_{\rm max}}$, $I_{m_{\rm max}}$ is $m_{\rm max}$-dimensional identity matrix, $m_{\rm max}$ is the maximum number of heavy atoms (CONF) that a molecule has for the dataset, $\phi$ is the activation function, and $W^\lambda$ is the matrix of weights for layer $\lambda \geq 1$, producing molecular descriptors illustrated in  Fig. \ref{fig:Fig1}.
The weight matrices are optimized by training the GCN using the algorithm illustrated in Fig. \ref{fig:Fig2}.  

The GCN are trained using the training and validation sets of molecules.  
For $\lambda=1$, $H^1 = X$ where $X$ are the node features in the one hot encoded data format. The feature of each node is a vector including atomic number, hybridization, degree, and chirality of each atom. Each node of the graph aggregates information from its neighbouring node(s) as the graph passes through layers of the GCN by Eq. (\ref{gcn_l+1}). The weight matrices are initialized randomly with a truncated normal distribution. The weight matrix of the last GCN layer is set as a one-dimensional vector, so that the resulting final layer $H^{\lambda+1}_{i}$ is a vector instead of a matrix. In this work, we build GCN with $3+1$ layers, where $W_1$ is a $20 \times 9$ matrix,  $W_2$ is a $9 \times 9$ matrix, and $W_3$ is a $9 \times 1$ matrix.
 %After multiple  layers, each node encodes with information not just of itself, but also information of nodes in proximity with it through chemical bonds.

The resulting $H^{\lambda+1}_{i}$ vector is treated as input into a Gaussian process regression (GPR) model, with entropy of the corresponding molecule as output. The RMSE over the validation set is used as the loss function for the optimization of the GCN weight matrices. 
The outputs of GCN are converted into a 9D vector by a densely connected NN layer, and the 9D vector is further reduced to $k$ dimensions by principal component analysis (PCA), yielding GCN-PC($k$) descriptors.

In the present work, the GCN model is trained with the cluster of constitutional isomers with the chemical formula C$_7$H$_{10}$O$_2$ from Ref. \cite{qm9_dataset}. The trained weight matrices are then used to compute the adjacency matrices for other molecules in the QM9 database of Ref. \cite{qm9_dataset} to generate the GCN-based descriptors. For smaller molecules with less than nine heavy atoms, the adjacency matrix size is still $9\times 9$.

%To expand on that work beyond alkanes, we demonstrated a similar result for molecules in the QM9 dataset, and that the correlation could be explained physically. Further explanation is in the discussion section (section \ref{sec:discussion}) .  

\subsection{Descriptor III: Gershgorin circle theorem}

\label{gershgorin}

The Gershgorin circle theorem identifies a region in the complex plane that contains all the eigenvalues of a complex square matrix \cite{Bellman+2021, saad2011numerical_gershgorin}. 
The theorem states that for an $m\times m$  matrix with the entries in complex plane $C$, the eigenvalues of matrix $M$ are in the range of $D_1$ $\cup$ $D_2$ $\cup$ ...$\cup$ $D_m$. Each disc $D_i$ is defined as
\begin{equation}\label{gc_disc_i}
    D_i = \{ z\in C: |{z-M_{ii}}|<\sum_{j\neq i}|M_{ij}|\}.
\end{equation}
Schrier \cite{schrier2020can} analyzed the eigenvalues of Coulomb matrices using the Gershgorin circle theorem and showed that a highly substituted carbon in an alkane has larger off-diagonal values, leading to higher upper bound for $D_i$ and consequently higher eigenvalues. We use the results of the Gershgorin circle theorem to define new descriptors of molecules, as follows. 

Each Gershgorin circle disc ($D_i$) of the coulomb matrix ($M$) is used to define a normal distribution function $f_i$
\begin{equation}
  f_i(x| \mu, \sigma^2) = \frac{1}{\sigma}\phi \left(\frac{x-\mu}{\sigma}\right),
\end{equation}
where $\mu$ and $\sigma$ are the mean and the standard deviation of the normal distribution, and $\phi(\cdot)$ is the probability density function (PDF) defined as 
\begin{equation}
  \phi (z) = \frac{e^{-z^2/2}}{\sqrt{2\pi}}.
\end{equation}
The mean of the normal distribution is defined with the value of the center of each Gershgorin circle disc, $M_{ii}$. The standard deviation of the distribution is determined by the radius of each Gershgorin circle disc, $(\sum_{j\neq i} |M_{ij}|)^\tau$. Each molecule is  then described by $f_{\rm molecule}$, defined as the sum of all $f_i$
\begin{equation}
  f_{\rm molecule} = \sum_i^{n_{\rm max}} d_i \times f_i(x| \mu = M_{ii}, \sigma^2 = (\sum_{j\neq i} |M_{ij}|)^\tau).
\label{gdes-1}
\end{equation}
where $d_i$ and $\tau$ are hyperparameters that allow adjustment on the weight on and standard deviation of $f_i$, respectively. In this work, we fix $d_i$ and $\tau$ to 1.  

We further define an atomic reference PDF ($f_{\rm atom}$), 
\begin{equation}
  f_{\rm atom} = d_{\rm atom} \times f_i(x| \mu = M_{\rm atom}, \sigma^2_{\rm atom}),
\label{gdes-1}
\end{equation}
where $M_{\rm atom} = 0.5 Z_{\rm atom}^{2.4}$, and $d_{\rm atom}$ is a weight constant that is set to 10. 
By calculating the inner products between the atomic reference PDF and the individual molecular PDF, we reduce the high dimensional continuous PDF curve into a five dimensional descriptor, $\langle f_{\rm HCNOF}, f_{\rm molecule} \rangle $. We also use the area under each $f_{\rm molecule}$ denoted hereafter by ${\rm AUC}(f_{\rm molecule})$ as an additional molecular descriptor. The relative role of these descriptors is illustrated in the following section. 

\begin{figure}[H]
\centering
\includegraphics[width=0.95\textwidth]{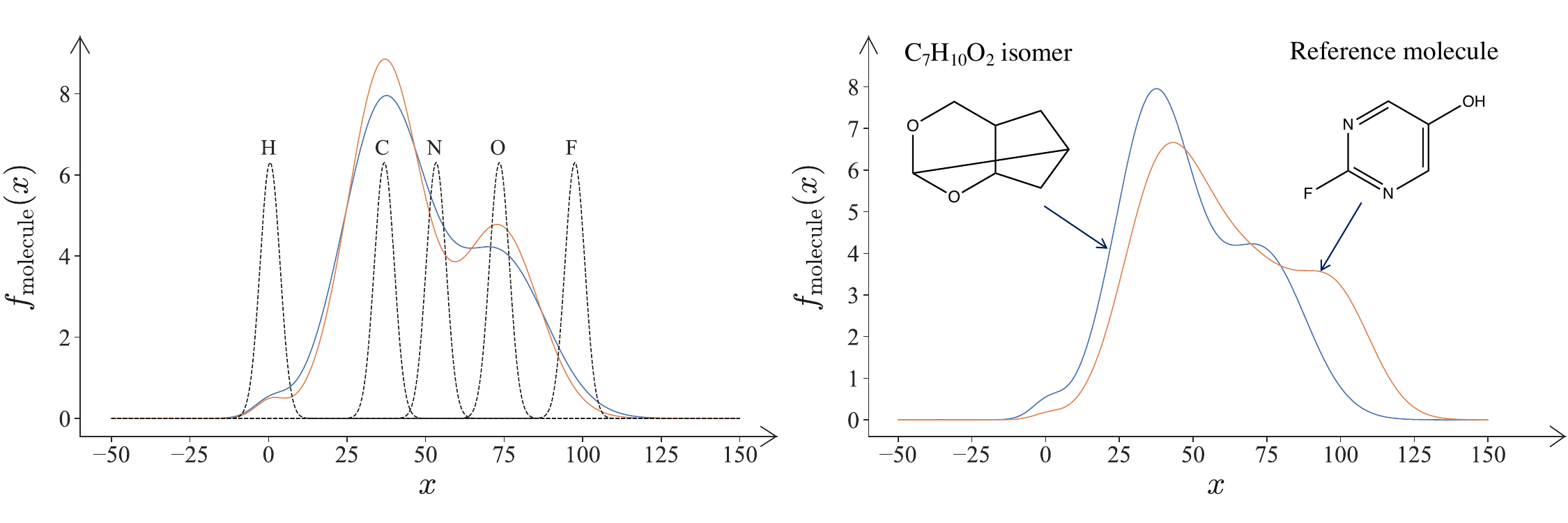}
\caption{\label{fig:Fig3} 
Left: Molecular probability density function for two isomers of ${\rm C_{7}H_{10}O_{2}}$  based on the Gershgorin circle theorem.  
The black doted curves show the atomic reference curves (HCNOF).
Right: Molecular probability density function for an isomer of ${\rm C_{7}H_{10}O_{2}}$ and the reference molecule based on the Gershgorin circle theorem.  For all $f_{\rm C_{7}H_{10}O_{2}}$ and $f_{\rm ref}$, $d_i = M_{ii}$, and $\tau = 1$.
}
\end{figure}

Figure \ref{fig:Fig3} shows  $f_{\rm{molecule}}$ of two $\rm  C_{7}H_{10}O_{2}$ constitutional isomers, where the blue curve is an isomer with higher entropy and the orange curve is that with lower entropy. The variances ($\sigma^2$) of $f_i$ are $\sum_{j\neq i} |M_{ij}|$. 
The atomic reference curves are shown by black dotted lines.

\subsection{ML method: Gaussian Process Regression with variable kernels}\label{GPR_model}
Gaussian process regression (GPR) \cite{murphy2018machine} is a supervised ML method that models an unknown target function, $f$, given input data $X$ and output data $y$. The training dataset comprises $ \mathcal{D} = \{(x_i,f_i), i \in [1,N_{\rm train}]\}$, where $f_i = f(x_i)$ is the noise-free observation of the function evaluated at $x_i$. GPR infers a distribution over functions with $p(f| \mathcal{D})$. In the context of property (e.g. entropy) interpolation across chemical space, a probability distribution of all possible entropy values for a molecule is inferred from the training dataset $\mathcal{D}$. The inputs $x_i$ are the molecular descriptors and $f_i$ is the corresponding value of molecular entropy. GPR defines a joint distribution to predict $f_*$ on the unseen inputs $(X_*)$ as
\begin{equation}\label{gp_definition}
    \begin{pmatrix}
    f \\
    f_{*}
    \end{pmatrix}\sim N\left(\begin{pmatrix}
    \mu \\
    \mu_*
    \end{pmatrix},\begin{pmatrix}
    K & K_* \\
    K_*^{T} & K_{**}
    \end{pmatrix}\right)
\end{equation}
where $\mu = \left [ m(x_1), ..., m(x_{N_{\rm train}}) \right ]$, with $m(\cdot)$ representing the mean of a Gaussian distribution, 
and $K=\kappa(X,X)$, $K_*=\kappa(X,X_*)$, $K_{**}=\kappa(X_*,X_*)$ determine the covariance matrix, defined by the kernel function $\kappa$. The efficiency of the GP model is determined by the functional form of the kernel function. 

In the present work, we follow Refs.\cite{duvenaud2011additive, duvenaud2013structure} to construct kernel functions with variable complexity by forming linear combinations and products of the following basis kernel functions:
\begin{equation}
    \kappa_{\rm{DP}}(x_i, x_j) = \sigma_0^2 + \langle x_i, x_j \rangle
\label{dp}
\end{equation}
where $\sigma_0$ is a parameter which controls the inhomogenity of the kernel,
\begin{equation}
    \kappa_{\rm{RBF}} (x_i, x_j) = \exp\left( - \frac{d(x_i, x_j)^2}{2l^2}\right)
\end{equation}
where $l$ is the length scale and $d(\cdot, \cdot)$ is the Euclidean distance between two data points, 
\begin{equation}
   \kappa_{\rm{RQ}}(x_i, x_j) = \left( 1 + \frac{d(x_i, x_j)^2}{2\alpha l^2} \right)^2
\end{equation}
where $\alpha$ is the scale mixture parameter and $l$ is the length scale of the kernel,
\begin{equation}
    \kappa_{\rm{MAT}}(x_i, x_j) = \frac{1}{\Gamma (\nu)2^{\nu -1}}\left( \frac{\sqrt{2\nu}}{l} d(x_i, x_j) \right)^{\nu} K_{\nu} \left(\frac{\sqrt{2\nu}}{l}d(x_i, x_j)\right)
\end{equation}
where $\nu$ and $l$ are positive parameters, $K_{\nu}(\cdot)$ is a modified Bessel function and $\Gamma(\cdot)$ is the gamma function, as well as %$d(\cdot, \cdot)$ is the Euclidean distance
\begin{equation}
    \kappa_{\rm{Periodic}}(x_i, x_j) = \exp\left(-\frac{2 \sin^2(\pi d(x_i, x_j))/p}{l^2}\right)
    \label{periodic}
\end{equation}
where $l$ is the length scale of the kernel, and $p$ is the periodicity of the kernel. % and $d(\cdot, \cdot)$ is the Euclidean distance.

\begin{figure}[H]
\centering
\includegraphics[width=1\textwidth]{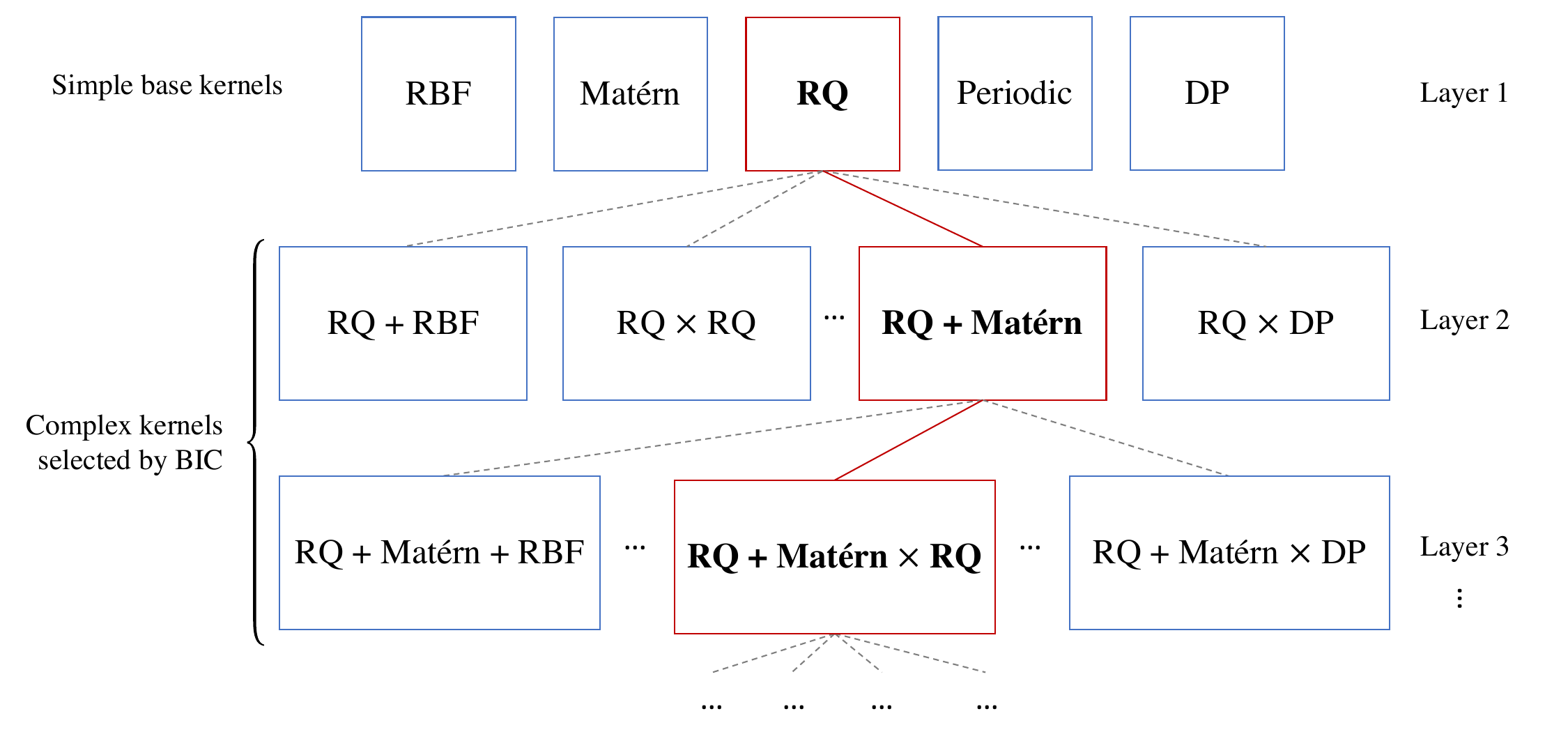}
\caption{\label{fig:Fig4} 
Schematic diagram of the kernel construction algorithm used in this work to increase the complexity of the kernel function for optimal kernel models. \\
}
\end{figure}

We optimize the functional form of the kernel function using a greedy search algorithm that combines simple functions into linear combinations and products, as was described in Refs. \cite{duvenaud2011additive, duvenaud2013structure} and used for various applications in Refs. \cite{torabian2023compositional, vargas2020physical, asnaashari2021gradient, dai2023neural, dai2020interpolation, vargas2018extrapolating}. The kernel construction algorithm is illustrated in Fig.  \ref{fig:Fig4}.

All possible linear combinations and products of base kernels defined in  Eqs. (\ref{dp}) - (\ref{periodic}) are created to yield more complex kernels. The model is selected by 
the Bayesian information criterion (BIC) calculated as
\begin{equation}
    {\rm BIC}(\mathcal{\kappa}_i) = \log P(y|{\kappa}_i) - \frac{1}{2}|{\kappa}_i| \log(N_{\rm train})
\end{equation}
where $|\kappa_i|$ is the number of parameters of kernel $\kappa_i$, and $\log P(y|{\kappa}_i) $ is the maximum of log-likelihood marginalized over the parameters of the functions in the Gaussian process, but not the kernel hyperparameters. This is the same function that is maximized to train a Gaussian process.

\section{Results}\label{sec:results}
\subsection{Analysis of different low-dimensional descriptors}

\begin{figure}[H]
\centering
\includegraphics[width=0.8
\textwidth]{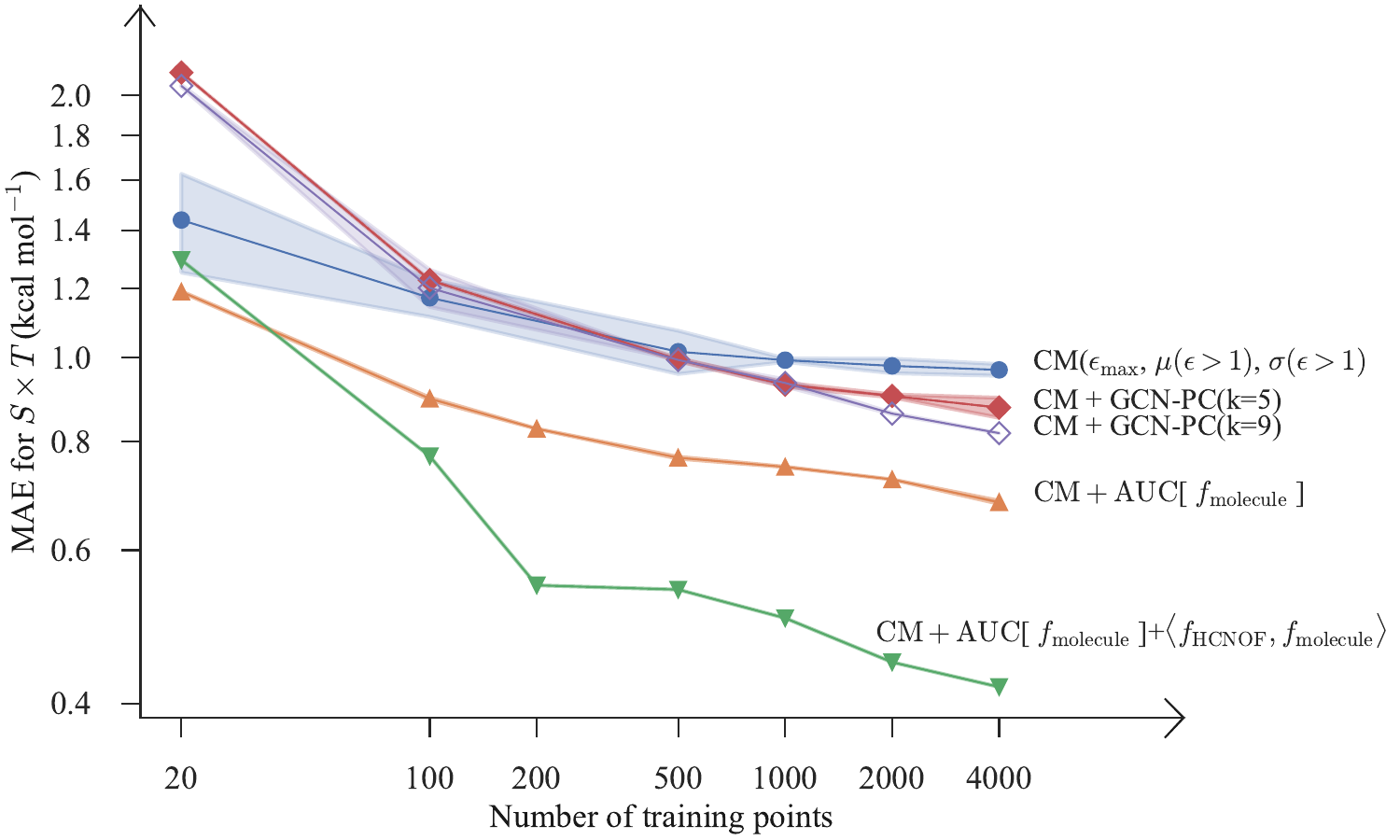}
\caption{\label{fig:Fig5} Test MAE of GP regression of entropy trained with different combinations of molecular descriptors: 
circles -- 3D CM($\epsilon_{\rm max}$, $\mu(\epsilon>1)$, $\sigma(\epsilon>1)$) with $\rm{RQ+RQ}$ kernel; filled-diamonds -- CM + GCN-PC($k = 5$)  with $\rm{RQ+RQ}$ kernel; empty diamonds -- CM + GCN-PC($k = 9$) with $\rm{RQ+RQ}$ kernel; up-triangles -- the 4D $\rm{CM}+\rm{AUC}[f_{\rm{molecule}}]$ with $\rm{RQ + RQ + MAT}$ kernel; down-triangles -- the 9D $\rm{CM}+\rm{AUC}[f_{\rm{molecule}}]+\langle f_{\rm HCNOF}, f_{\rm molecule} \rangle$ descriptor trained with $\rm{RQ + MAT \times DP}$ kernel.
The kernel of each model is optimized as described in text with BIC as the model selection metric. 
 The lines represent the average of ten sets of training using different distributions of molecules in the training set, and the shaded area is the standard deviation of the results. 
 }
\end{figure}

In order to analyze the relative performance of the different molecular descriptors, we begin by building GP models of entropy with various combinations of molecular descriptors described in the previous section.
%The GP models are trained as described.
Figure \ref{fig:Fig5} compares the learning curves of GP models thus obtained.
As shown in Fig. \ref{fig:Fig5}, the best performing GP model is obtained with nine dimensional CM-GC descriptor, which includes three descriptors stemming from the Coulomb matrix denoted as CM($\epsilon_{\rm max}$, $\mu(\epsilon>1)$, $\sigma(\epsilon>1)$), the area under the curve ${\rm AUC}(f_{\rm molecule})$, and five inner products $\langle f_{\rm HCNOF}, f_{\rm molecule} \rangle$.   The figure shows that models trained with CM-GC descriptor yield the most accurate interpolation, reducing the MAE to below 1 $\rm{kcal~mol^{-1}}$ with less than 100 molecules. This figure illustrates that physical descriptors derived from combinations of CM-based quantities and properties based on the Gershgorin circle theorem improve the accuracy of ensuing models much more significantly than the descriptors based on graph neural networks.

\begin{figure}[H]
\centering
\includegraphics[width=1\textwidth]{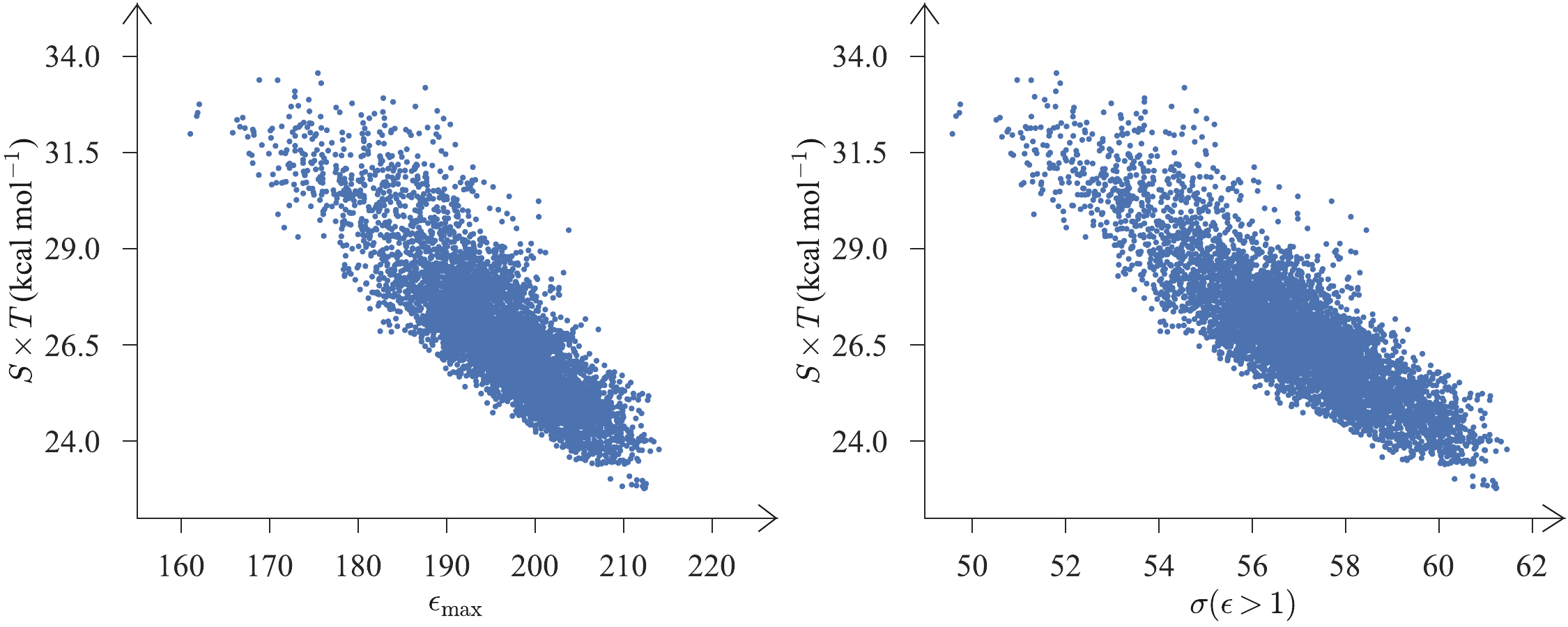}

\caption{\label{fig:Fig6} Entropy as a function of the largest eigenvalue ($\epsilon_{\rm max}$) of the Coulomb matrix (left panel) and the standard deviation ($\sigma $) of the Coulomb matrix eigenvalues greater than 1 (right panel) for constitutional isomers with the formula C$_7$H$_{10}$O$_2$.
}
\end{figure}

To understand the role of the descriptors contributing to the best performing combination in Figure \ref{fig:Fig5}, 
we examine correlations of molecular entropy with the largest eigenvalue of the Coulomb matrix and the standard deviation $\sigma$ of the distribution of the eigenvalues $\epsilon > 1$, both shown in Fig. (\ref{fig:Fig6}). While neither $\epsilon_{\rm max}$ nor $\mu$ are unique features for entropy, the results in Figs. \ref{fig:Fig6} illustrate that there exist strong correlations between entropy and both of $\epsilon_{\rm max}$ and $\sigma$. These correlation can be rationalized using the Gershgorin circle theorem.

The Gershgorin circle theorem identifies a region in the complex plane that contains all the eigenvalues of a complex square matrix \cite{Bellman+2021, saad2011numerical_gershgorin}. 
The theorem states that for an $m\times m$  matrix with the entries in complex plane $C$, the eigenvalues of matrix $M$ are in the range of $D_1$ $\cup$ $D_2$ $\cup$ ...$\cup$ $D_m$. Each disc $D_i$ is defined in Eq. \ref{gc_disc_i}.
Schrier \cite{schrier2020can} analyzed the eigenvalues of Coulomb matrices using the Gershgorin circle theorem and showed that a highly substituted carbon in an alkane has larger off-diagonal values, leading to higher upper bound for $D_i$ and consequently higher eigenvalues. 

\begin{figure}[H]
\centering
\includegraphics[width=0.8\textwidth]{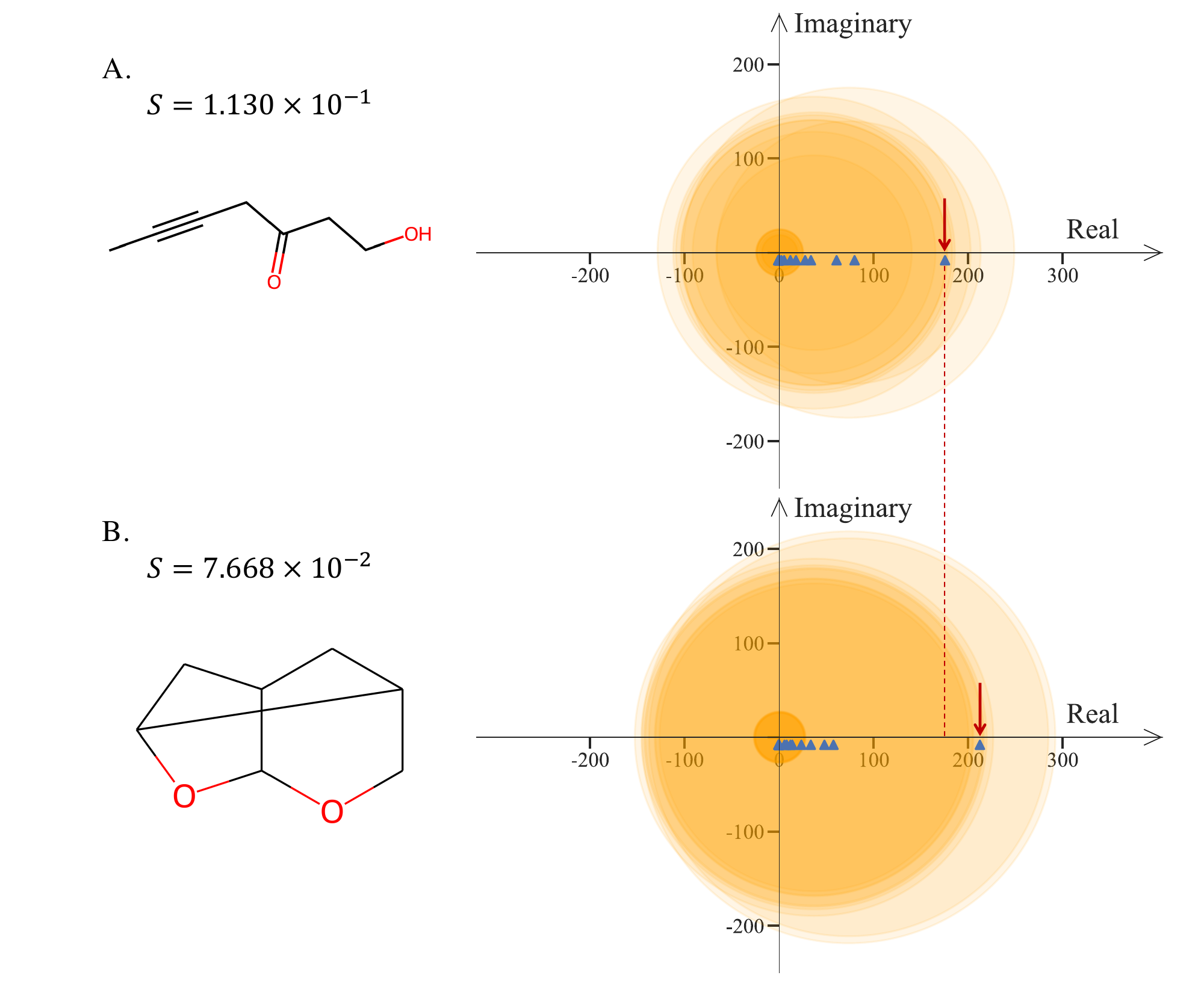}
\caption{\label{fig:Fig7} Visual demonstration of the Gershgorin circle theorem on the Coulomb matrices of two C$_7$H$_{10}$O$_2$ constitutional isomers 1-hydroxyhept-5-yn-3-one ({\bf A}, top) and 2,4-dioxatricyclo[$4.3.0.0^{3,8}$]nonane ({\bf B}, bottom), where the molecule {\bf A} has the higher entropy. The entropy values are in the unit of kcal mol$^{-1}$.The Gershgorin discs of each isomer's Coulomb matrix are drawn as orange circles in the figures. The blue triangles are all the eigenvalues of the Coulomb matrices. The red arrows emphasis the CM-EV1 of each isomer's Coulomb matrix.}
\end{figure}

To elucidate the connection with entropy, consider two constitutional isomers with molecular formula C$_7$H$_{10}$O$_2$  that have drastically different entropy. 
The molecules with the highest and lowest entropy in the C$_7$H$_{10}$O$_2$ isomeric cluster are 1-hydroxyhept-5-yn-3-one (Fig. \ref{fig:Fig7}.A) and 2,4-dioxatricyclo[$4.3.0.0^{3,8}$]nonane (Fig. \ref{fig:Fig7}.B), respectively. For molecule {\bf A}: 
the entropy is $1.130\times 10^{-1}$ kcal mol$^{-1}$K$^{-1}$,  $\epsilon_{\rm max} = 175.437$, $\sigma(\epsilon>1)= 51.800$ and $\mu(\epsilon>1) = 45.987$. For molecule {\bf B}: 
the entropy is $7.668\times 10^{-2}$ kcal mol$^{-1}$K$^{-1}$ (lower),  $\epsilon_{\rm max} = 212.320$ (higher), $\sigma(\epsilon>1) = 61.221$ (higher), while $\mu(\epsilon>1) = 45.980$ is similar to that of 1-hydroxyhept-5-yn-3-one. Molecule {\bf B} is more rigid, which leads to fewer microstates and lower entropy.  
\\

\begin{table}[H]
\begin{center}
  \begin{tabular}{p{2cm}p{2cm}p{2cm}p{4.5cm}p{1.5cm}}%{|p{2cm}|p{2cm}|p{2cm}|p{5cm}|p{2cm}|}%{  l  l  l  l  l  }
    \hline % \hline
    Atom 1 & Atom 2 & $Z_1 Z_2$ & $R$~(\AA) & $M_{ij}$\\ \hline  \hline
    H & H & 1 & 0.74 & 1.351 \\ \hline
    C & H & 6 & 1.09 & 5.505 \\ \hline
    O & H & 8 & 0.971 & 8.239 \\ \hline
    C & C & 36 & 1.535 (single) & 23.453\\ 
     & &  & 1.339 (double) & 26.886 \\ 
     & &  & 1.203 (triple) & 29.925 \\ \hline
    C & O & 48 & 1.431 (single, ethanol) & \textbf{33.543} \\ 
     &  &  & 1.219 (double, ketone) & \textbf{39.377} \\ 
    \hline %\hline
  \end{tabular}
\caption{Off-diagonal entries ($M_{ij}$) of the Coulomb matrix for several bonded atoms.} 
\label{table:Mij_values}
\end{center}
\end{table}

The more rigid molecule, 2,4-dioxatricyclo [$4.3.0.0^{3,8}$]nonane (Fig. \ref{fig:Fig7}B.), has a lower entropy. This is the manifestation of the fact that entropy is determined by the shape and rigidity of the molecule \cite{huyskens1989influence}. Rigid molecules tend to be restricted in motion, which leads to lower numbers of available microstates and lower entropy. Thus, a good representation of molecular rigidity should be a good descriptor for predicting entropy. 
Given this premise, it is useful to analyze how the CM($\epsilon_{\rm max}$, $\mu(\epsilon>1)$, $\sigma(\epsilon>1)$) descriptor reflects the rigidity of molecules. 
%CM($\epsilon_{max}$, $\mu$, $\sigma$)
 The diagonal entries of CM  (Table \ref{table:Mij_values}) for molecules considered here are $M_{ii,{\rm H}} = 0.5$,  $M_{ii,{\rm C}} = 36.858$, or $M_{ii,{\rm O}} = 73.517$, corresponding to hydrogen, carbon and oxygen. The off-diagonal elements $M_{ij}$ are directly proportional to the atomic numbers $Z_i$ and $Z_j$, and inversely proportional to the interatomic separation $R_{ij}$. This means that the $M_{ij}$ value is larger for bonded pairs of heavy atoms. Table \ref{table:Mij_values} lists some examples of $M_{ij}$ values, where the C$=$O double bond pair gives the highest $M_{ij}$ value. For rigid molecules, the atoms are spatially crowded, which leads to higher $M_{ij}$ values on average. We observe that the Gershgorin circle theorem yields higher values of $|A_{ij}|$ in Eq. (\ref{gc_disc_i}) for rigid molecules on average, leading to larger values of $\sum_{j\neq i}|A_{ij}|$. Thus, the range of eigenvalues defined by the union of $D_i$ in the circle theorem is higher for rigid molecules. The opposite is true for linear and less rigid molecules.

Fig. \ref{fig:Fig7} visualizes the results of the Gershgorin circle theorem. The $y$- and $x$-axes are the imaginary and real axes of the complex plane containing the eigenvalues. The orange discs are $D_i$ defined by Eq. (\ref{gc_disc_i}). Fig. \ref{fig:Fig7} shows that the union of discs for the more rigid molecule on the right is $D_1 \cup ...\cup D_{19} = \{ -152.77 \leq z \leq 292.16\}$, which covers a wider area compared to that for the molecule on the left, $D_1 \cup ...\cup D_{19} = \{ -128.78 \leq z \leq 248.76\}$. Since the range of eigenvalues is wider for the more rigid molecule, it has a higher probability of having a larger $\epsilon_{\rm max}$ value and bigger $\sigma(\epsilon>1)$. Since the molecules from the same constitutional isomer cluster have the same atomic composition, the distribution of the center of the discs, ${D_i, i \in N}$, are identical. This explains the similarity of $\mu(\epsilon>1)$ for molecules from the same cluster observed in our analysis (not shown) of the dataset.

\begin{figure}[H]
\centering
\includegraphics[width=1\textwidth]{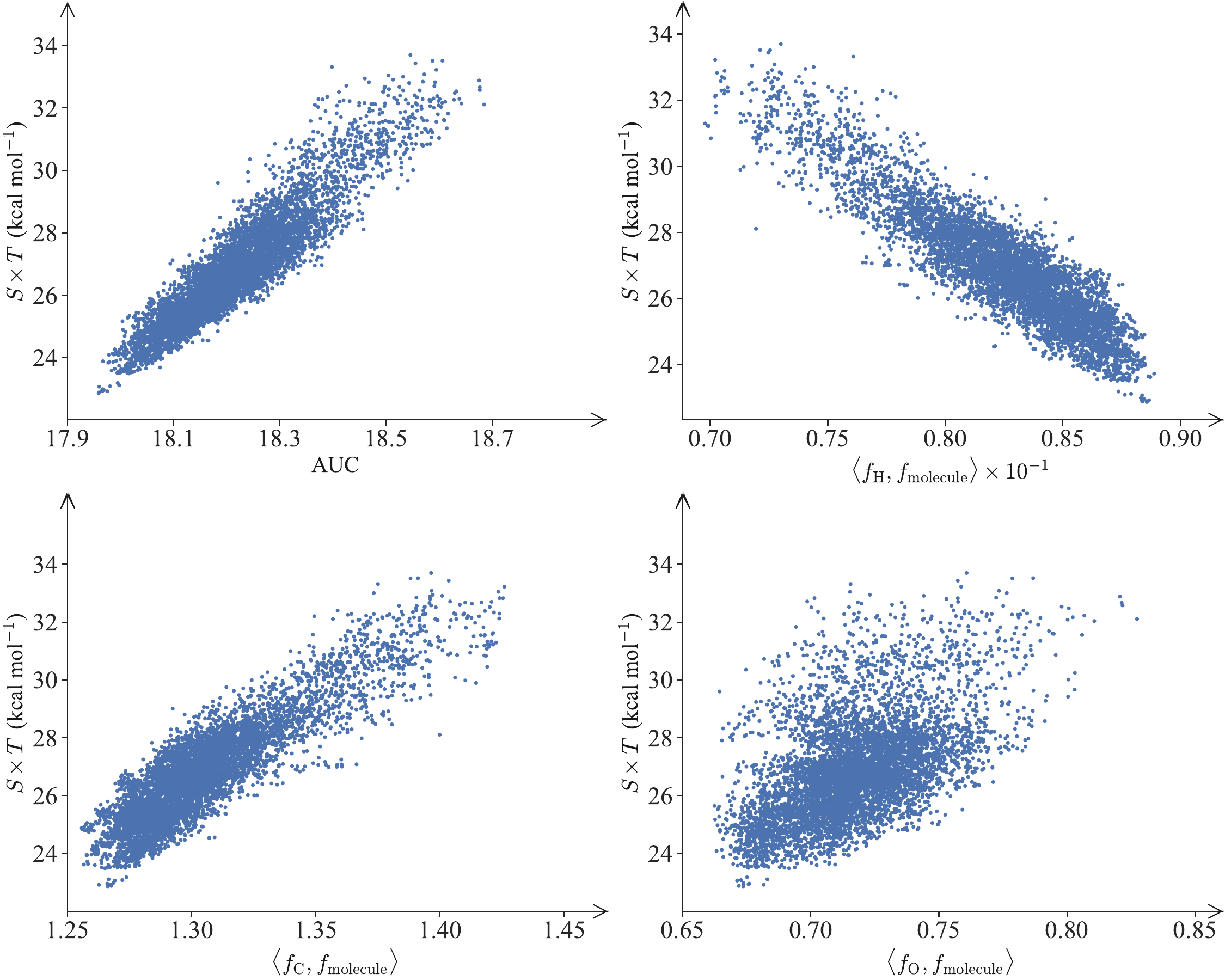}
\caption{\label{fig:Fig8} Correlation of entropy with AUC (upper left panel) and $\langle f_{\rm atom}, f_{\rm molecule} \rangle$ for constitutional isomers with the chemical formula $\rm C_7H_{10}O_2$. 
 } 
\end{figure}

To illustrate the physical significance of the descriptors based on PDF derived from the Gershgorin circle theorem as defined by Eq. (\ref{gdes-1}), we examine in Figure \ref{fig:Fig8} the correlation of entropy with each of the individual overlaps $\langle f_{i}, f_{\rm molecule} \rangle$ and with ${\rm AUC}(f_{\rm molecule}(x| \mu = M_{ii}, \sigma^2 = (\sum_{j\neq i}|M_{ij}|)^2 )$.
Figure \ref{fig:Fig8} shows that entropy is particularly well correlated with $\langle f_{\rm H}, f_{\rm molecule} \rangle$ and $\langle f_{\rm C}, f_{\rm molecule} \rangle$
and strongly correlated with ${\rm AUC}(f_{\rm molecule}(x| \mu = M_{ii}, \sigma^2 = (\sum_{j\neq i}|M_{ij}|)^2 )$. 

\subsection{Data-efficient models of entropy}

\begin{figure}[H]
\centering
\includegraphics[width=1\textwidth]{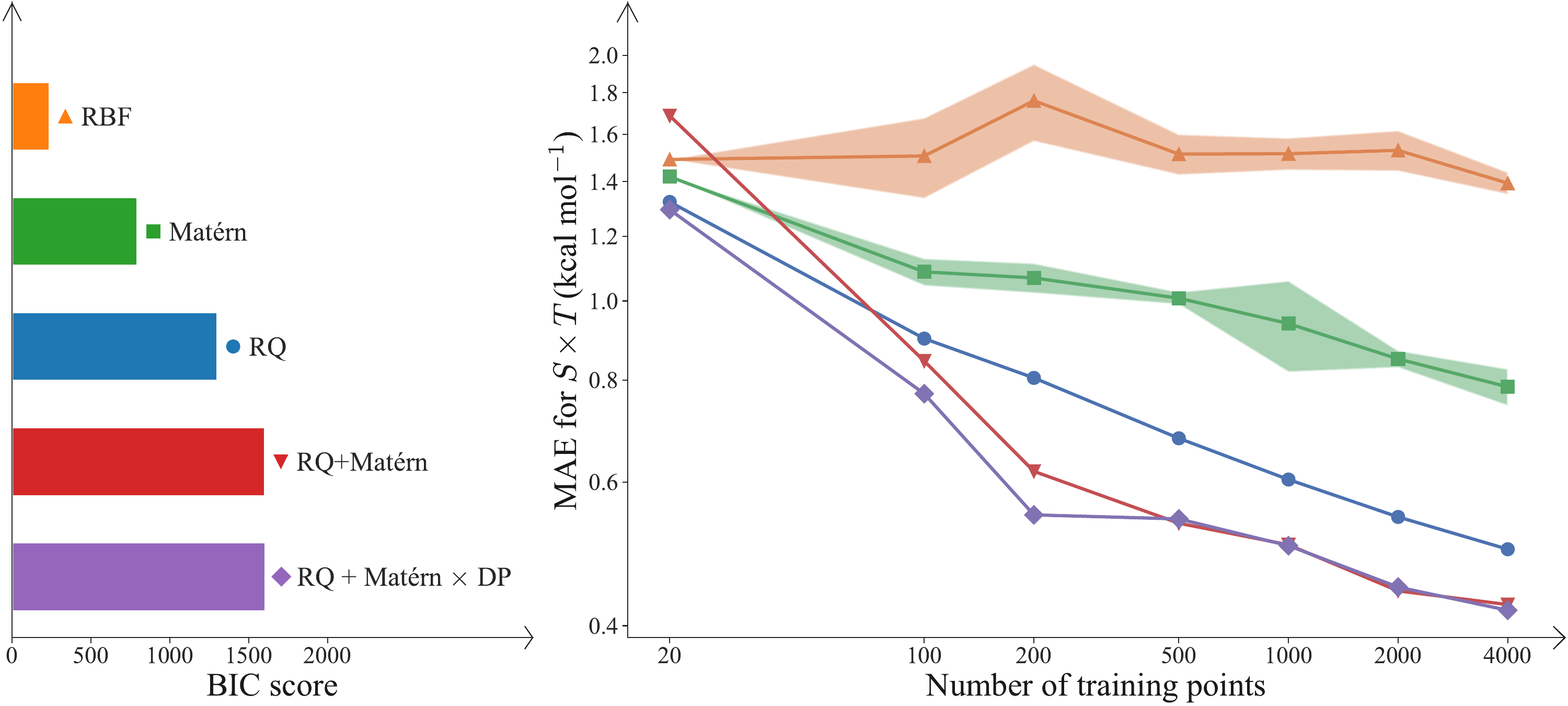}
\caption{\label{fig:Fig9} Left: BIC values for models with different kernels. Right: Learning curves of GP models  trained with the same 9D descriptor ($\rm{CM}+\rm{AUC}[f_{\rm{molecule}}]+\langle f_{\rm HCNOF}, f_{\rm molecule} \rangle$), but with the different kernels illustrated in the left panel. The shaded area is the standard deviation computed with ten models trained with randomly sampled data points. For the lines not showing the shaded area, the standard deviation is too small to be visible on this scale. 
}
\end{figure}

All results in the present work show entropy multiplied by temperature at $T = 298.15$~K in the units of kcal mol$^{-1}$. Our goal is to build models with prediction accuracy below 1 kcal mol$^{-1}$ using as few molecules on input as possible. 
We aim to achieve this by exploiting the most performant descriptors identified in the previous section in combination with kernel models based on optimized kernels. When the dimensionality of descriptors is low, it is possible to build anisotropic kernels, optimized for each dimension of the input space separately. In the present work, we have found that including the anisotropy into the kernels for the best low-dimensional descriptors of molecules does not have any significant effect on the accuracy of the ensuing models.  

We use GP regression to model entropy. The kernel functions of the GP models are chosen and optimized using greedy search in the space of kernel functions guided by BIC as discussed in the ML method section.  
This aligns kernels with available data as much as possible, given the constraints of the  greedy search, the descriptors and the uncertainties stemming from the extremely limited size of the datasets considered here. To reduce bias in the calculation of BIC used for the greedy search of the optimal kernel function, 1000 training data points are chosen randomly from the pool of ~85,000 data points reserved for training. Each of the randomly chosen training datasets is passed through multiple layers of the BIC kernel selection algorithm. At each level of kernel selection, we calculate the average value of BIC scores, $\mu$(BIC), for the same kernel trained with ten different sets of randomly selected data points. The base or complex kernel with the highest average BIC score is retained for the next level of kernel function selection.

The effect of the kernel function on the accuracy of the resulting models of entropy is shown in Fig. \ref{fig:Fig9}. 
Figure \ref{fig:Fig9} compares the $\mu$(BIC) values of base and complex kernels. Although the $\rm{RQ} + \rm {Matern} \times \rm{DP}$ kernel has the highest BIC score, this BIC score is only slightly higher than that of the $\rm{RQ}+\rm{Matern}$ kernel. The $\rm{RQ}+\rm{Matern}$ kernel is therefore chosen as the optimal kernel for the best model of entropy in this work due to its high BIC score and lower function complexity. The BIC algorithm stops at level four because the BIC scores plateau and decrease with further increase of the kernel complexity.

\subsection{Zero-point Vibrational Energy}

\begin{figure}[H]
\centering
\includegraphics[width=1\textwidth]{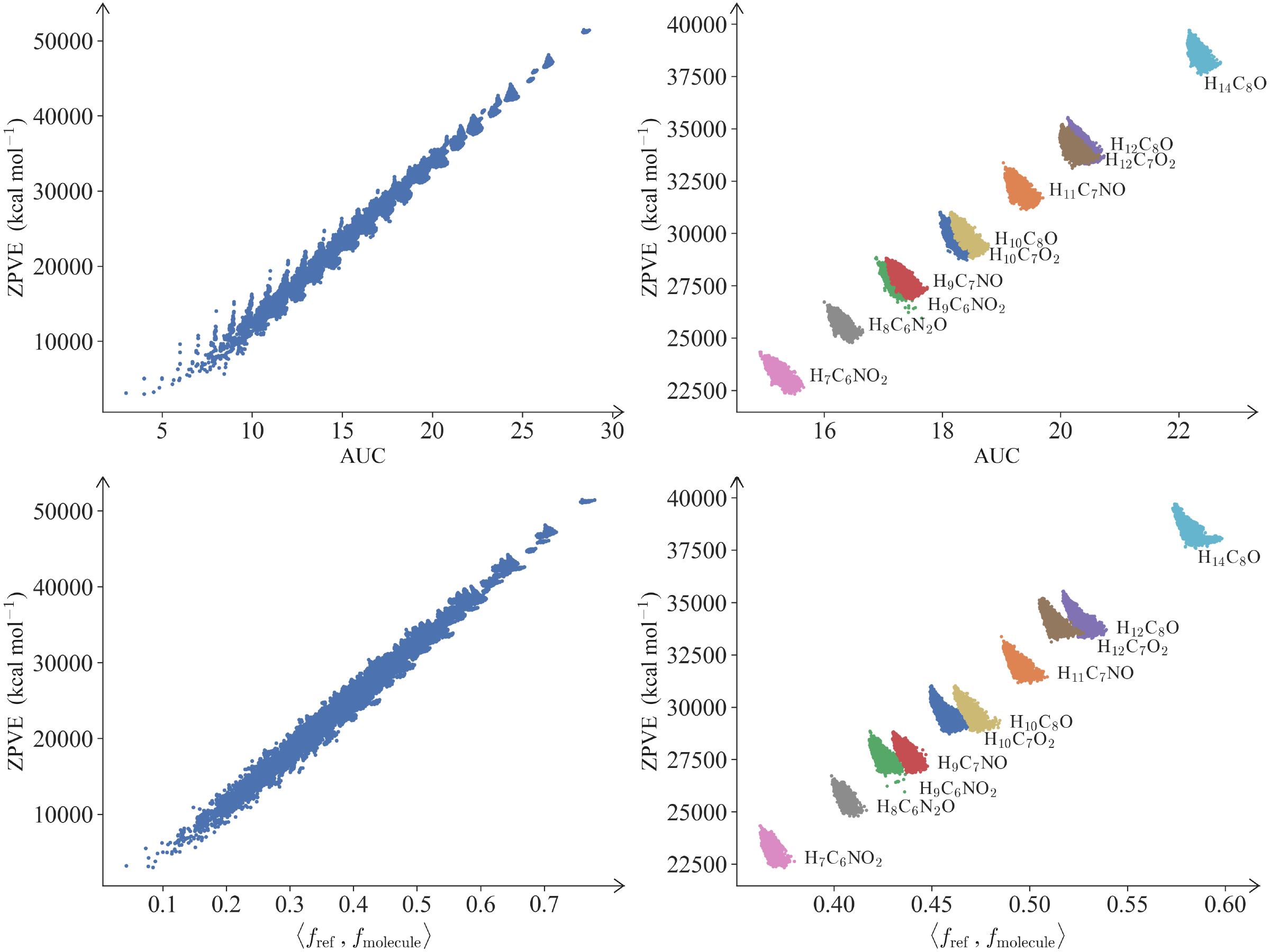}
\caption{\label{fig:Fig10} Zero point vibrational energy as a function of ${\rm AUC}(f_{\rm molecule}(x| \mu = M_{ii}, \sigma^2 = (\sum_{j\neq i}|M_{ij}|)^2 )$ and $\langle f_{\rm ref}, f_{\rm molecule} \rangle$ for all molecules in the QM9 dataset (left panels); and the molecules from ten largest isomer groups in the QM9 dataset (right panels). The reference molecule is ${\rm H_3C_4N_2OF}$ illustrated in Fig. \ref{fig:Fig3}. 
 }
\end{figure}

The preceding sections illustrate that very efficient and accurate models of entropy can be built with 9D descriptors ($\rm{CM}+\rm{AUC}[f_{\rm{molecule}}]+\langle f_{\rm HCNOF}, f_{\rm molecule} \rangle$).  
It is instructive to explore the performance of these descriptors in models of other molecular properties. Here, we consider zero point vibrational energy (ZPVE) of molecules. 
Figure \ref{fig:Fig10} illustrates that ZPVE is strongly correlated with ${\rm AUC}(f_{\rm molecule})$ and $\langle f_{\rm ref}, f_{\rm molecule} \rangle$.

\newpage

\begin{figure}[H]
\centering
\includegraphics[width=0.8\textwidth]{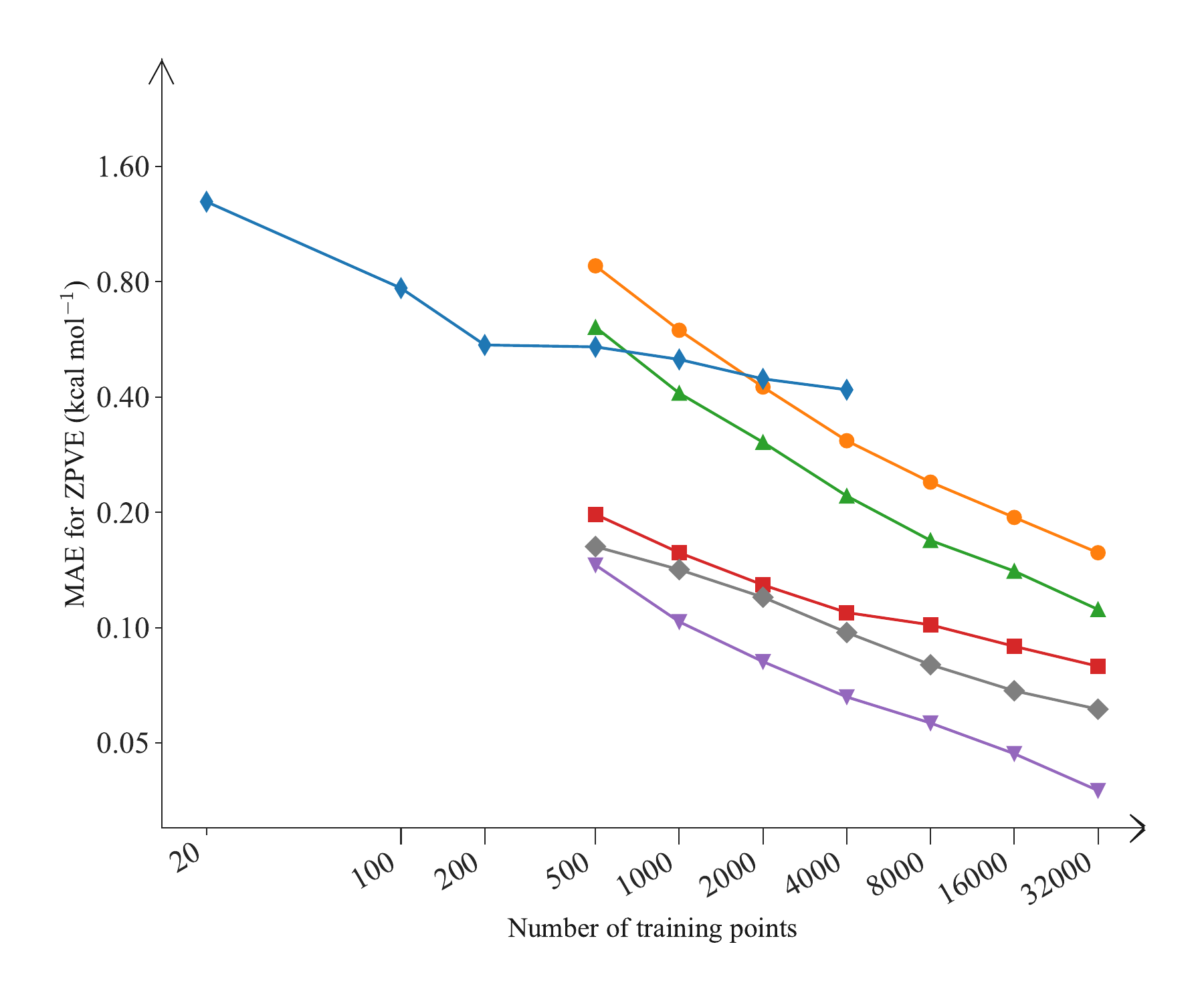}
\caption{\label{fig:Fig11} Comparison of the present results (squares) of ZPVE with the results of Lilienfeld et al. (2023) \cite{khan2023quantum}. The present results are obtained 
with 9D molecular descriptors ${\rm CM} + {\rm AUC} + f_{\rm molecule} + \langle f_{\rm HCNOF}, f{\rm molecule}\rangle$ and the GP model with the optimized ${\rm RQ+ MAT \times DP}$ kernel. The results of Ref.  \cite{khan2023quantum} are based on kernel ridge regression with the following descriptors: full CM including 435 features  (orange circles), BOB with 1128 features (green triangles), SLATM 11960 features (red squares), FCHL with 10440 features (purple down triangles), and MBDF with 145 features (brown diamonds). 
}
\end{figure}

Figure \ref{fig:Fig11} shows the interpolation MAE for ZPVE illustrating the comparison of the performance of the present 9D models with models developed in Ref. \cite{khan2023quantum} . 
The results in Ref. \cite{khan2023quantum} use high-dimensional descriptors:  full CM including 435 features  (shown in Fig. \ref{fig:Fig11} orange circles), BOB with 1128 features (shown by green triangles), SLATM 11960 features (shown by red squares), FCHL with 10440 features (shown by purple down triangles), and MBDF with 145 features (shown by brown diamonds). These high-dimensional descriptors were used in Ref. \cite{khan2023quantum} to compute simple isotropic kernels for kernel ridge regression models of ZPVE.

\begin{figure}[H]
\centering
\includegraphics[width=0.8\textwidth]{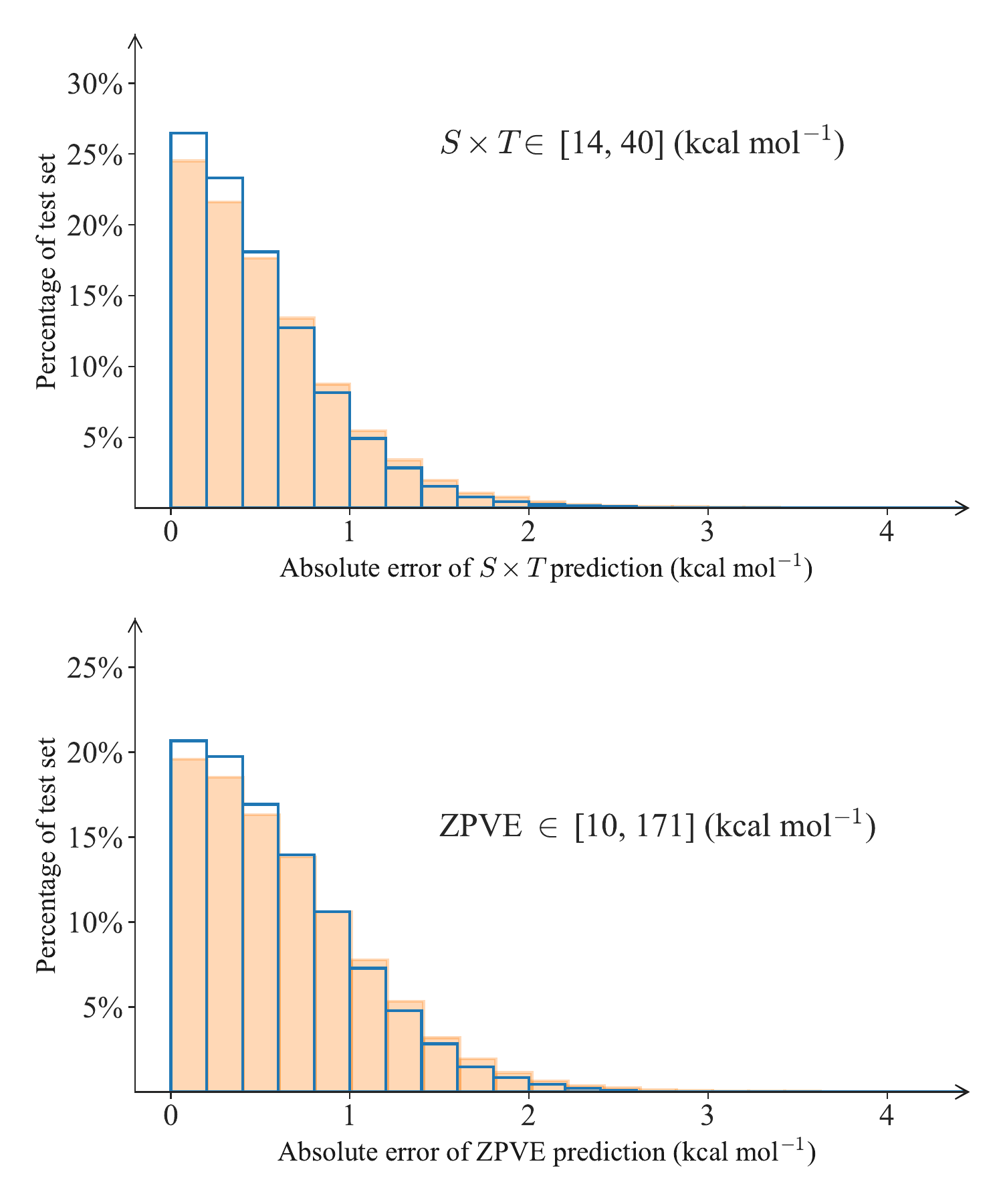}
\caption{\label{fig:Fig12} Distribution of errors for predictions of entropy (upper) and ZPVE (lower) with nine-dimensional descriptors of molecules identified as the best descriptors in Fig. \ref{fig:Fig5}. Full bars -- predictions of models trained with 100 molecules; open bars -- models trained with 1000 molecules.}
\end{figure}

Our results shown in Fig. \ref{fig:Fig11} by blue diamonds use nine-dimensional descriptors with optimized, complex kernels given by  the ${\rm RQ+ MAT \times DP}$ combination of base kernels. These kernels are used for GP regression models of ZPVE. The results illustrate that the 9D descriptors proposed here produce ZPVE interpolation models with the prediction MAE under 1 kcal mol$^{-1}$, when trained by as few as 100 molecules.

While it is common to quantify the model performance by average errors, such metrics do not illustrate fully the quality of the models. A combination of average errors and largest errors is a better metric, which, however, can be significantly affected by a small number of outliers. To quantify non-ambiguously the performance of the bets models of entropy and ZPVE developed in the present work, we show in Fig. \ref{fig:Fig12} the distribution of errors of predictions over a hold-out test set, containing 20,000 molecules of different size. The shaded bars illustrate the performance of models trained with 100 molecules and the open bars -- molecules trained with 1000 molecules. We note that the range of $S \times T$ for the molecules considered here covers 26 kcal mol$^{-1}$, while the range of ZPVE covers 160 kcal mol$^{-1}$. In both cases, we observe that our models trained with 100 molecules are able to predict $S \times T$ and ZPVE with absolute error under 1 kcal mol$^{-1}$ for $> 78$ \% of the test data and with error under 1.3  kcal mol$^{-1}$ for $> 92.4$ \% of the test data.

\section{Summary}

The goal of the present work is to build efficient, data-starved models of molecular properties for interpolation in chemical compound spaces comprising molecules of variable complexity with low-dimensional molecular descriptors. The dataset considered here comprises 133,885 molecules with complexity ranging from 3 atoms to 29 atoms. We consider two molecular properties: entropy and zero point vibrational energy. We begin by designing low-dimensional descriptors tailored for the prediction of entropy. Our results show that entropy can be efficiently interpolated by nine-dimensional models with molecular descriptors comprising three physical parameters describing the distributions of eigenvalues of Coulomb matrices and six parameters accounting for the composition and shape of molecules through quantities based on the Gershgorin circle theorem. 

In order to build efficient interpolation models of entropy, we combine the nine-dimensional descriptors thus obtained with Gaussian process regression based on kernels with variable functional form. The kernel functions are optimized using a greedy search algorithm guided by the Bayesian information criterion (BIC). In order to obtain meaningful, unbiased models based on extremely restricted datasets, we use BIC averaged over multiple randomly sampled instances of training data as a model selection metric.  Nine-dimensional Gaussian process models thus optimized are capable of predicting molecular entropy across the entire dataset with the average error under 1~kcal mol$^{-1}$, using as few as 100 molecules on input. We present a careful analysis of model prediction errors, demonstrating that entropy models trained with 100 molecules predict $S \times T$ with chemical accuracy under 1~kcal mol$^{-1}$ for $> 80$ \% of molecules in the test set, comprising 20,000 molecular species. 

The nine-dimensional descriptors based on the distributions of the Coulomb matrix eigenvalues and the Gershgorin circle theorem are then shown to perform well for the prediction of a completely different molecular property - ZPVE. The error analysis illustrates that the Gaussian process models of ZPVE trained with 100 molecules and using optimized kernels predict ZPVE with absolute error under 1 kcal mol$^{-1}$ for $> 80$~\% of the test set and with error under 1.3 kcal mol$^{-1}$ for $> 90$ \% of molecules in the test set. This is remarkable, given that ZPVE of molecules in the test set covers the range from 10 to 171 kcal mol$^{-1}$. 

There are three main implications of the results presented here. First, because the training sets of the demonstrated models are very small, accurate predictions of molecular properties for a wide range of molecules can be based on either experimental data or extremely accurate quantum chemistry calculations. Second, the possibility of using low-dimensional descriptors for accurate predictions with restricted datasets opens up (1) the possibility of building models with anisotropic kernels that account for the difference in scales of different features; and (2) the possibility of Bayesian optimization in chemical compound space, which can be used to predict and design molecules with desired properties. We will explore both of these possibilities in a future work.

\section*{Acknowldgments}

This work was supported by NSERC of Canada.

% \bibliographystyle{jabbrv_asprev4-1}
%\bibliographystyle{jabbrv_asprev4-1} 
%\bibliographystyle{jabbrv_abbrv}
%\bibliography{bibliography}
\bibliographystyle{iopart-num}
\bibliography{bibliography_ordered}

\end{document}